\documentclass[pra,superscriptaddress,a4paper,twocolumn,showpacs]{revtex4}
\usepackage{amsmath,amsfonts, graphicx,color}
\def\>{\rangle}
\def\<{\langle}
\def\E{ {\cal E} }
\def\oE{ \overline{{\cal E}} }
\def\F{ {\cal F} }
\def\D{ {\cal D} }

\def\I{ I }
\def\orho{\overline{\rho}}
\def\Tr{ \mbox{Tr} }
\def\non{ \nonumber\\}
\def\plus{ |+\> }
\def\minus{|-\> }

\begin{document}

\title{Quantum computation via measurements on the low temperature state of a many-body system} %
\author{David Jennings}%
\affiliation{School of Physics, The University of Sydney, Sydney, New South Wales 2006, Australia}%
\affiliation{Institute for Mathematical Sciences, Imperial College London, London SW7 2BW, United Kingdom}%
\author{Andrzej Dragan}%
\affiliation{Optics Section, Blackett Laboratory, Imperial College London, London SW7 2BZ, United Kingdom}%
\affiliation{Institute of Theoretical Physics, University of Warsaw, Ho\.{z}a 69, 00-681 Warsaw, Poland}%
\author{Sean D. Barrett}%
\affiliation{Optics Section, Blackett Laboratory, Imperial College London, London SW7 2BZ, United Kingdom}%
\affiliation{Institute for Mathematical Sciences, Imperial College London, London SW7 2BW, United Kingdom}%
\affiliation{Research Centre for Quantum Science and Technology,
Macquarie University, NSW 2109, Australia}%
\author{Stephen D. Bartlett}%
\affiliation{School of Physics, The University of Sydney, Sydney, New South Wales 2006, Australia}%
\author{Terry Rudolph}%
\affiliation{Optics Section, Blackett Laboratory, Imperial College London, London SW7 2BZ, United Kingdom}%
\affiliation{Institute for Mathematical Sciences, Imperial College London, London SW7 2BW, United Kingdom}%

\date{23 September 2009}

\begin{abstract}
We consider measurement-based quantum computation using the state of
a spin-lattice system in equilibrium with a thermal bath and free to
evolve under its own Hamiltonian. Any single qubit measurements
disturb the system from equilibrium and, with adaptive measurements
performed at a finite rate, the resulting dynamics reduces the
fidelity of the computation. We show that it is possible to describe
the loss in fidelity by a single quantum operation on the encoded
quantum state that is independent of the measurement history. To
achieve this simple description, we choose a particular form of
spin-boson coupling to describe the interaction with the
environment, and perform measurements periodically at a natural rate
determined by the energy gap of the system.  We found that an
optimal cooling exists, which is a trade-off between keeping the
system cool enough that the resource state remains close to the
ground state, but also isolated enough that the cooling does not
strongly interfere with the dynamics of the computation. For a
sufficiently low temperature we obtain a fault-tolerant threshold
for the couplings to the environment.
\end{abstract}

\maketitle

\section{Introduction}

The ``one-way'' model for quantum computation, which requires only
local adaptive measurements of individual qubits prepared in a fixed
multi-qubit resource state, provides a new approach for assessing
the physical requirements for universal quantum computing. The
\emph{cluster state} on a two-dimensional square lattice is the
canonical example of a resource state that allows for universal
measurement-based quantum computation (MBQC)~\cite{Rau01,Rau03}.
Much research on MBQC focusses on properties of the resource state
itself, and in particular how such a state could be prepared
dynamically via, say, local controlled-$Z$ operations in a variety
of systems for which the dynamics of the individual qubits can be
uncoupled, such as an optical lattice~\cite{Tre06} or single
photons~\cite{Bro05,Wal05}.  In contrast, recent new theoretical
results in MBQC have been obtained by viewing the resource state as
the ground or low-temperature thermal state of a strongly-coupled
quantum many-body system~\cite{Rau05,Bar06, Bre08, Doh08, Barr08,
Skr09}. This perspective allows us to use some powerful tools and
techniques of quantum many-body theory, for example, to determine
what type of systems permit universal MBQC~\cite{Gross08,
Gross07,Bre09} and for those that do, how robust the system is in
its universality~\cite{Rau05,Doh08, Barr08,Skr09}.

One could take this perspective of ground states serving as
computational resources as a physical realisation, and thus obtain a
new mechanism for creating cluster states or other such resource
states.  That is, if a quantum many-body system could be engineered
such that it possesses the cluster state as its unique ground
state~\cite{Bar06}, and if the system is sufficiently gapped, then
MBQC can be performed by cooling the system down to a sufficiently
low temperature and then performing a sequence of adaptive
measurements on this thermal state.  However, by treating the
resource state for MBQC as the equilibrium state of a dynamical
system, any measurements that we perform will necessarily disturb it
from its thermal state. As measurements must be adaptive, if they
are separated by finite time intervals we are faced with both errors
produced by the evolution under the system's Hamiltonian and also
the cooling interaction with the environment. These two sources of
dynamical error, together with thermal errors, act to reduce the
output fidelity of any MBQC scheme that we might wish to perform.

In this paper, we investigate a regular lattice of qubits for which
the free Hamiltonian has the cluster state as its ground state. The
system is first prepared by cooling via a simple and convenient
choice of coupling to a bosonic bath in a thermal state, and we
assume that the coupling to the bath is present throughout the
computation. This situation is relevant to an experiment in which a
strongly coupled system is first prepared in a useful initial
resource state using a refrigerator, and which cannot easily be
subsequently decoupled from the refrigerator before the MBQC
commences. Alternatively, in the context of a laser-cooled atomic
system, it may be inconvenient or undesirable to turn off the
cooling interaction before the MBQC commences.

We explicitly determine how MBQC proceeds on this system's thermal
state as it is perturbed by measurements, with free evolution and
cooling interaction ongoing between measurements. In particular, for
the lattice of spins in the presence of a spin-boson coupling to a
thermal bath that acts to restore the system to the pure cluster
state, we show that the free Hamiltonian for the spin lattice state
induces coherent oscillations that determine a natural measurement
rate. Importantly, the effect of the bath may be conveniently
described by a single quantum operation that acts on the encoded
quantum information within the MBQC computation and which is
independent of the particular measurement history. With this result
we demonstrate that MBQC on such a dynamical thermal state is
fault-tolerant for a sufficiently low temperatures and for couplings
to the bath below a given threshold.

\section{MBQC with Dynamics}

The cluster state on a lattice $\mathcal{L}$ of $N$ qubits may be
defined in the stabilizer formalism~\cite{Nie00} as the unique
eigenstate of each of the mutually commuting stabilizers $K_i =
X_i\otimes_{j \sim i}Z_j$ with eigenvalue one~\cite{Rau03}. Here $i$
labels a particular site in the lattice and $j \sim i$ signifies
that $j$ is a neighbouring site of $i$. The stabilizer description
allows us to define the cluster state as the ground-state of the
Hamiltonian
\begin{equation}\label{hamc}
  H_c = -\frac{\Delta}{2} \sum_{i\in\mathcal{L}} K_i\,,
\end{equation}
with an energy gap $\Delta $. A state in the $k^{\mbox{\tiny th}}$
excited level of this Hamiltonian is obtained by performing $Z$
errors at $k$ distinct sites, and implies that the energy level is
$\binom{N}{k}$-fold degenerate.

A useful alternative description of the cluster state, which we
shall make use of shortly, is in terms of the action of entangling
unitaries between neighbouring qubits on the lattice. The qubits at
all sites are first initialized in the state $|+\>$, i.e.,
stabilized by the set of operators $X_i$, and then for every bond in
the lattice the controlled-$Z$ unitary $\exp(i\pi |1\rangle\langle
1| \otimes |1\rangle \langle 1|)$ is performed between the two end
qubits. We denote the product of all controlled-$Z$ operations on
each bond simply as $U$, and the link between the two descriptions
is provided by the relation $U K_i U^\dagger = X_i$.

MBQC on the ground state of a spin lattice model governed by this
Hamiltonian involves an adaptive measurement procedure, in which
qubits are measured sequentially in different bases until the
desired output state is produced, up to Pauli operator corrections,
on the remaining unmeasured qubits.  The $Z$ measurements play a
special role and are used to remove individual qubits from the
cluster state, while sequences of single qubit measurements in the
$X$-$Y$ plane are used to implement unitary gates on the encoded
quantum information. For the perfect cluster state, the inputs can
be taken, without loss of generality, to be $|+\>$ on each of the
qubits to be measured first.

\subsection{Measurements and free evolution of the lattice}

As some of these measurements are adaptive (i.e., the choice of
measurement bases are conditional on prior measurement outcomes)
they must necessarily be performed at different times.  Measurements
disturb the system out of its ground state, and between measurements
this disturbed state will evolve under the Hamiltonian (\ref{hamc}).
An important property of this Hamiltonian is that it is
dispersionless, and so any localized excitations will remain local
and will not propagate across the lattice. For example, if a $Z$
measurement is performed at site $s$ the system is projected into an
equal superposition of the ground-state $U(\otimes_i |+ \>_i)$ and
the state with a single $Z$ error on site $s$, $U(\otimes_{ i\ne
s}|+ \>_i \otimes | - \>_s)$. In the case of a single $X$
measurement at the site $s$, the system is projected into a
superposition of the ground-state $U(\otimes_i |+ \>_i)$ and the
state with $Z$ errors on all of the neighbouring sites of $s$,
$U(\otimes_{i\not\sim s} |+ \>_i \otimes_{i \sim s} |-\>_i)$. This
local disturbance remains local under evolution; however, because
the post-measurement state is no longer an energy eigenstate, this
evolution must be accounted for when we perform subsequent
measurements on neighbouring qubits.

If all of the measurements on the system can be performed on a
timescale much less than that of the system's evolution, one may be
able to treat the effect of short-time evolution as a small
perturbation of the cluster state.  However, one could alternatively
make use of a natural timescale of this system. The equal-spaced
spectrum of the Hamiltonian (\ref{hamc}), with spacing $\Delta$,
ensures that evolution is periodic with period $\tau = 2\pi/\Delta$.
If (instantaneous) measurements are made at time intervals which are
multiples of this period, the evolution of the system under the
Hamiltonian can be ignored.  In essence, the energy gap $\Delta$ of
the Hamiltonian provides a natural ``clock speed'' for the quantum
computation.  Given that the gap in the system will determine the
temperature to which the lattice must be cooled in order to approach
the ground state, it will be desirable to engineer systems in which
this gap is as large as possible; with this in mind, performing very
fast measurements (i.e., at a frequency $\gg \Delta$) may not be
possible, and performing measurements at this clock speed (or
integer fractions thereof) is a much less stringent requirement.

\subsection{Spin-boson model}

The quantum many-body system with Hamiltonian (\ref{hamc}) is
gapped, and so we can prepare a cluster state (or a close
approximation to it) by cooling the system down to near its ground
state though coupling to a thermal bath. (This cooling can be done
efficiently because of the simplicity of the
Hamiltonian~\cite{Ver08}.)  Performing measurements on the ground
state yields excited states that are no longer in equilibrium with
the bath and so, if the cooling interaction is present, any
measurement scheme that we may perform on the cluster state must
proceed sufficiently quickly to avoid a return to equilibrium.
However, we have already argued that the free Hamiltonian $H_c$ will
require measurements to be close to the intervals $2\pi/\Delta$, and
this clock speed provides a lower bound on the overall duration of
the computation.  We now consider the effect of a finite measurement
rate in the presence of such cooling.

To model the effects of cooling, we consider a system
consisting of a bath of bosons held at a low temperature and coupled
to each site qubit via a spin-boson interaction, which takes the
form $H_I = \sum_{i,j} \lambda_{ij} Z_i q_{ij}$.  Here,
$q_{ij}=a_{ij}+a^\dagger_{ij}$ is the displacement operator for the
$j^{\mbox{\tiny th}}$ mode at site $i$ and $\lambda_{ij}$ are
coupling constants. The full Hamiltonian for the system of qubits
and bosons is then
\begin{equation}
  \label{ham}
  H_{\mbox{\tiny tot}} = H_c + H_I + H_b\,,
\end{equation}
where $H_b = \sum_{ij} \omega_{ij} a^\dagger_{ij} a_{ij}$ is the
free Hamiltonian for the bath.

We note that our results depend on this choice of $Z$ axis to
describe the coupling to the bath. In practice we may not have full
control over this coupling, but in many systems (e.g., in trapped
atoms), to a good approximation the environment couples only to a
single spin component of the qubit degree of freedom.  In such a
situation, one may take the coupling to the bath as defining the $Z$
axis.  This assumes we have full control over the cluster
Hamiltonian, and so may adjust it so as to coincide the $Z$ axis for
the cluster state with the axis defined by the cooling interaction.
We leave as open the question of how MBQC can proceed with a more
general coupling.

Because the interaction Hamiltonian commutes with the set of
controlled-$Z$ unitaries applied to every neighbouring pair of
qubits, we can map this system using the unitary $U$ to a dual
system of uncoupled qubits, with the same interaction $H_I$ to the
thermal bath.  We shall consider the master equation for this dual
system with total Hamiltonian
\begin{equation}
  \overline{H}_{\mbox{\tiny tot}} \equiv
  UH_{\mbox{\tiny tot}}U^\dagger
  = -\frac{\Delta }{2} \sum_i X_i + H_I + H_b\,.
\end{equation}
In general, we denote operators in the dual model with a overline
bar. For example, single-qubit measurements on the original system,
given by projectors $P_k$, are now described in this dual model by
multi-qubit projectors $\overline{P_k} = UP_k U^\dag$.

A standard derivation results in a master equation~\cite{walls95}
for the lattice subsystem given by
\begin{multline}
  \label{master}
  \dot{ \orho}(t) =  \sum_i \Bigl( i \frac{\Delta }{2} [X_i, \orho (t)] +
  \alpha_i \D \bigl[ |+\>_i \<-| \bigr]\orho (t) \\
  +\beta _i \D \bigl[ |-\>_i \<+| \bigr]\orho (t) \Bigr)\,,
\end{multline}
where the action of the superoperator $\D[A]$ on the state $\orho $
is given by $\D[A] \orho =A \orho A^\dagger - \frac{1}{2} \{ A^\dagger
A, \orho \}$. The constants $\alpha_i$ and $\beta _i$ are parameters
that depend on the couplings to the bath, $\lambda_{ij}$, and the
temperature of the bath, $T$.  They are given explicitly as
\begin{align}\label{couplings}
 \alpha_i &= 2 \pi \sum_j \lambda _{ij}^2 \delta (\omega _{ij} - \Delta ) (1+ n (\Delta )) \non
 \beta_i  &= 2 \pi \sum_j \lambda _{ij}^2 \delta (\omega _{ij} - \Delta ) n (\Delta ) \non
 n(E) &= (e^{E/kT} -1)^{-1} \,.
\end{align}
We make the simplifying assumption that the couplings $\alpha$ and
$\beta$ do not vary from site to site and for later reference we may relate the temperature of the bath to the coupling parameters through the equation
\begin{eqnarray}\label{temperature}
kT &=& \Delta / (\log (\alpha/\beta)).
\end{eqnarray}

Within the dual picture, a qubit initially in the state $\orho_0$
will evolve in time $t$ under a completely-positive (CP) map $\oE_t$
to the state $\orho_t = \oE_t(\orho_0)$.  A Kraus decomposition for
this CP map $\oE(\orho ) = \sum_i M_{i,t} \orho M_{i,t}^\dagger$, is
given by
\begin{align}
  \label{kraus}
  M_{1,t} &= \sqrt{\frac{\alpha}{\alpha +\beta}}\bigl(e^{-i \Delta t} |+ \> \< +| + e^{-(\alpha +\beta) t/2} |- \> \<- |\bigr) \non
  M_{2,t} &= \sqrt{\frac{\beta}{\alpha +\beta}} \bigl(e^{+i \Delta t} |- \> \< -| + e^{-(\alpha +\beta) t/2} |+ \> \<+ |\bigr) \non
  M_{3,t} &= \sqrt{\alpha(1-e^{- (\alpha +\beta) t})/(\alpha+\beta)} |+ \> \<- | \non
  M_{4,t} &= \sqrt{\beta (1-e^{- (\alpha +\beta) t})/(\alpha +\beta)} |- \> \<+ |\,.
\end{align}
This evolution takes any single qubit state $\orho$  asymptotically
in time towards an equilibrium state
\begin{eqnarray}
 \orho_e&=&\frac{\alpha }{\alpha + \beta } |+ \>\< +| + \frac{\beta }{\alpha + \beta } |- \>\< -|\non
  &=&\frac{1 }{1 + e^{-\Delta/kT} } |+ \>\< +| + \frac{e^{-\Delta/kT} }{1 + e^{-\Delta/kT} } |- \>\< -|\,.
  \label{eq:Equilibriumrho}
\end{eqnarray}
Thermal equilibrium for the full lattice is reached with a rate
governed by the couplings $\alpha$ and $\beta$.

\subsection{Example: Arbitrary X-rotation}
\label{subsec:Xrot}

To illustrate the effect of dynamics on MBQC we consider performing
a simple single-qubit gate using MBQC on a one-dimensional lattice.
More general gates will behave similarly, as we shall show in
Sec.~\ref{sec:General}.

Consider performing an arbitrary $X$-rotation gate, i.e., a rotation
$X(\theta) = \exp[-i\frac{\theta}{2} X]$ of a single qubit about the
$X$-axis by angle $\theta$. The smallest cluster state that can
realise such a gate is the three-qubit cluster state on a line.  The
ideal gate proceeds as follows for a non-dynamical cluster state.
The qubits are initially prepared in the state $|\psi_{\rm
in}\>_1\otimes \plus _2 \otimes \plus _3$. The state is then
entangled with the unitary $U$. Qubit 1 is measured in the basis $\{
\plus , \minus \}$, with measurement result $s_1 \in \{0,1\}$. Based
on this measurement result, qubit 2 is measured in the basis
$\{\exp(-i\frac{\eta }{2} Z) \plus, \exp(-i\frac{\eta }{2} Z)
\minus\}$, where $\eta = (-1)^{s_1}\theta$, with measurement result
$s_2$. For the static case it is simple to show that, subsequent to
these measurements, qubit 3 is left in the state $ \exp[-i\frac{\eta
}{2}X_3]Z_3^{s_1}X_3^{s_2}|\psi_{\rm} \>_3 =
Z_3^{s_1}X_3^{s_2}\exp[-i\frac{\theta}{2}X_3]|\psi_{\rm} \>_3$. That
is, the initial state $|\psi_{\rm} \>$ has been rotated by the gate
$\exp[-i\frac{\theta}{2}X]$ up to Pauli operator corrections
$Z_3^{s_1}X_3^{s_2}$.

For a dynamical three-qubit cluster state that evolves according to
the Hamiltonian (\ref{ham}), the timing of the two projective
measurements becomes important for the gate to succeed with high
fidelity.  First, if the initial state is left to interact with the
bath, the system would eventually evolve to the equilibrium state
and the input state $\orho_{\mbox{\tiny in}} = |\psi_{\mbox{\tiny
in}}\rangle \langle \psi_{\mbox{\tiny in}}|$ would be erased. For
temperature $T\ge 0$ we assume that the initial state is
$U(\orho_{\mbox{\tiny in}}\otimes \orho_e^{\otimes 2}) U^\dagger$,
where $\orho_e$ is given by Eq.~(\ref{eq:Equilibriumrho}), and that
the system evolves for a time $t_0$ until the projective measurement
$P_1$ on qubit 1; the measured state then evolves until time $t_0
+t$ at which point the second measurement $P_2$ is performed. The
output state is thus given by
\begin{equation}\label{evolved}
  \rho = U \overline{P}_2 \oE_t (\overline{P}_1\oE_{t_0} (\orho_{\rm in}
  \otimes \orho_e ^{\otimes 2})\overline{P}_1^\dagger)\overline{P}_2^\dagger
  U^\dagger\,,
\end{equation}
where we described the evolution in our dual model, related to our
system by the unitary operation $U$, with evolution $\oE_{t_i}$ at
time $t_i$ obtained from (\ref{master}), and $\overline{P_k} = U P_k
U^\dagger$.

The evolution up to time $t_0$ is given by
\begin{equation}
  \oE_{t_0}: \orho _{\mbox{\tiny in}}\otimes \orho_e^{\otimes 2}
  \mapsto \oE_{t_0}(\orho _{\mbox{\tiny in}})\otimes \orho_e^{\otimes 2}\,.
\end{equation}
For convenience, we define $\orho_{t_0} = \oE_{t_0}(\orho _{\mbox{\tiny
in}})$, which can be expressed in Bloch vector form as $\orho_{t_0}
=\frac{1}{2} (\I + \vec{r}_{t_0}\cdot \vec{\sigma})$ with
$\vec{r}_{t_0}=(x_{t_0},y_{t_0},z_{t_0})$.

\begin{figure}
\includegraphics[width=7.5cm ]{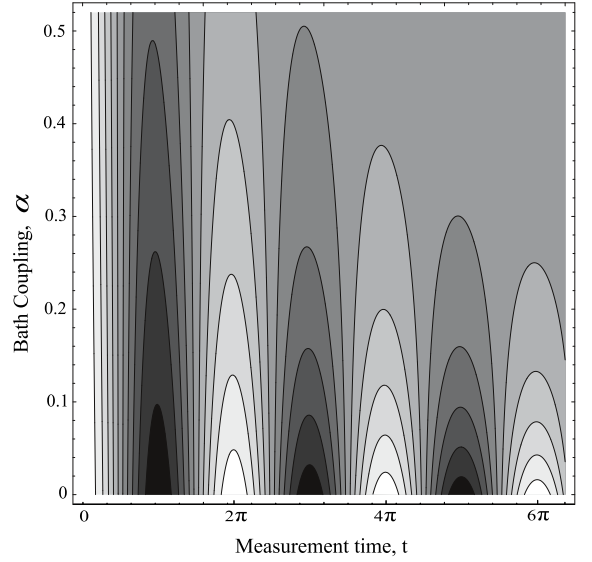}
\caption{Contour plot of fidelity as a function of coupling
$\alpha $, to a zero temperature bath and measurement time $t$. We set $\Delta =1$  and show ten equally
spaced contours between $F=0$ $(\alpha =0, t = \pi )$ and $F=1$
$(\alpha =0, t = 2\pi n )$. Each shaded band corresponds to an interval of 10\%, for example, the white regions correspond to a fidelity of
$90\le F \le 100\%$, centered around multiples of $\tau$, the uppermost
large grey region corresponds to $40 \le F \le 50 \%$ while the black regions correspond to $0 \le F \le 10\%$.
}\label{contour}
\end{figure}

The second stage of the evolution is different due to the projective
measurement on the first qubit. A direct calculation of
(\ref{evolved}) followed by tracing out qubits 1 and 2 yields the
final state of qubit 3
\begin{equation}
 \orho_3(t_0,t) = Z_3^{s_1}X_3^{s_2} e^{-i \frac{\theta}{2} X_3}
 \orho_{\rm \tiny out} e^{i\frac{\theta }{ 2}X_3} X_3^{s_2}Z_3^{s_1} \,,
\end{equation}
where $\orho_{\rm \tiny out} = \frac{1}{2}[ \I +\vec{r}_{\rm
\tiny out}(t_0,t)\cdot \vec{\sigma}]$ and
\begin{align}\label{rout}
\vec{r}_{\rm out}(t_0,t) &= (x_{\mbox{\tiny out}},y_{\mbox{\tiny
out}},z_{\mbox{\tiny out}})\non
 x_{\mbox{\tiny out}} &= x_{t_0}e^{-\frac{1}{2} (\alpha +\beta) t}\cos  \Delta t\non
 y_{\mbox{\tiny out}} &= (y_{t_0}\cos  \Delta t - z_{t_0}\sin  \Delta t)e^{-( \alpha +\beta) t}\cos \Delta t\non
 z_{\mbox{\tiny out}} &= (z_{t_0}\cos  \Delta t + y_{t_0}\sin  \Delta t)e^{-\frac{3}{2}(\alpha +\beta) t}\,.
\end{align}
The decoherence due to the evolution under the coupling to the bath
does not depend on the particular choice of unitary that we perform,
and furthermore the fidelity, being unitarily invariant, depends
only on $\vec{r}_{\mbox{\tiny out}}$. For the situation of a perfect
cluster state ($T=0$) with $t_0=0$ and $\orho _{\mbox{\tiny in}}=|+
\> \< +|$ we have
\begin{equation}
  \label{fidel}
  F(\orho _{\mbox{\tiny in}}, \orho _{\mbox{\tiny out}} )
  =\frac{1}{2} (1 + e^{-\frac{1}{2} \alpha  t} \cos  \Delta t)\,,
\end{equation}
which is plotted in Fig.~\ref{contour}.  For a fixed $\alpha$, the
local maxima for fidelity occur slightly before the times given by
multiples of $\tau = 2\pi/\Delta$, due to the presence of the
exponential factor, however we note that the analysis derived from
the master equation (\ref{master}) will only be valid for weak bath
couplings $\alpha \ll \Delta $.

We see that to obtain high fidelity, we perform the measurements at
times given by multiples of $\tau$. In the large $t$ limit the
output state decoheres to the maximally mixed state $\frac{1}{2}
\I$, which reflects a return to the pure cluster state. We also note
that the evolution of the encoded quantum information between time
$0$ and time $t_0$ is distinct from the evolution between $t_0$ and
$t_0+t$ and we will show that, in general, the latter form of
evolution is the typical way in which fidelity is lost. For
comparing results here with those obtained in the general
decoherence situation, we note that the measurement on qubit 1
produces a Hadamard transformation of the encoded state, and
consequently swaps the $x$ and $z$ components of the Bloch vector.

\subsection{Optimal Cooling Rate}\label{cooling}

%

In any experimental realization of MBQC on a strongly coupled system, there will be a residual thermal coupling to an ambient background (the environment), at temperature $T_{\mbox{\tiny bg}}$. Typically, this environment is warm compared to the relevant energy scale in the system, i.e.  $kT_{\mbox{\tiny bg}} \gtrsim \Delta$. The coupling to this background can be reduced, for example by screening the system from thermal noise, but usually it can not be eliminated altogether. The purpose of the cooling bath (at temperature $T_{\mbox{\tiny bath}}$) is to counteract this residual heating effect, by actively cooling the system such that the lattice of spins is prepared in a highly entangled cluster state at an effective temperature $kT \ll \Delta, kT_{\mbox{\tiny bg}}$.
%
%
However, as described in the previous section, the coupling to this bath also has an unwanted effect, which is to reduce the fidelity of MBQC on the system, by disrupting the state of the system over the course of the computation. A reasonable question to ask, then, is how the fidelity of a calculation varies as the strength of the coupling to the cooling bath is varied.

The effects of a cooling bath plus high-temperature background may
be modeled by including separate Lindblad terms for each of the baths in the
master equation:
\begin{multline}
  \label{twobathsfull}
  \dot{ \orho}(t) =  \sum_i  i \frac{\Delta}{2} [X_i, \orho (t)]  \\
  +  \sum_i \Bigl(\alpha_{\mbox{\tiny bath}} \D [ |+\>_i \<-| ]\orho (t)
  + \beta_{\mbox{\tiny bath}} \D [ |-\>_i \<+| ]\orho (t) \Bigr)\\
  +  \sum_i \Bigl(\alpha_{\mbox{\tiny bg}} \D [ |+\>_i \<-| ]\orho (t)
  + \beta_{\mbox{\tiny bg}} \D [ |-\>_i \<+| ]\orho (t) \Bigr)\, ,
\end{multline}
where $\alpha_{\mbox{\tiny bath}}$ and $\beta_{\mbox{\tiny bath}}$ describe the coupling to the cooling bath at temperature $T_{\mbox{\tiny bath}}$, and $\alpha_{\mbox{\tiny bg}}$ and $\beta_{\mbox{\tiny bg}}$ are the corresponding coupling strengths to the background environment at temperature $T_{\mbox{\tiny bg}}$. For simplicity, we assume that the background is very warm compared to the energy gap in the system, $kT_{\mbox{\tiny bg}} \gg \Delta$, and use (\ref{temperature}) to deduce that  $\alpha_{\mbox{\tiny bath}} = \beta_{\mbox{\tiny bath}} \equiv \gamma$, and also that the cooling bath is at a very low temperature $kT_{\mbox{\tiny bath}} \ll \Delta$, such that  $\beta_{\mbox{\tiny bath}}=0$. In this limit, the master equation becomes
\begin{multline}
  \label{twobaths}
  \dot{ \orho}(t) =  \sum_i  i \frac{\Delta}{2} [X_i, \orho (t)]
  +  \sum_i \alpha_{\mbox{\tiny bath}} \D [ |+\>_i \<-| ]\orho (t)
   \\
  +  \gamma \sum_i \Bigl( \D [ |+\>_i \<-| ]\orho (t)
  +\D [ |-\>_i \<+| ]\orho (t) \Bigr)\, .
\end{multline}
(Note that the effect of a non-zero temperature cooling bath can also be described by this master equation by a suitable redefinition of $\alpha_{\mbox{\tiny bath}} $, $\beta_{\mbox{\tiny bath}} $ and $\gamma$).

\begin{figure}
\includegraphics[width=8.5cm ]{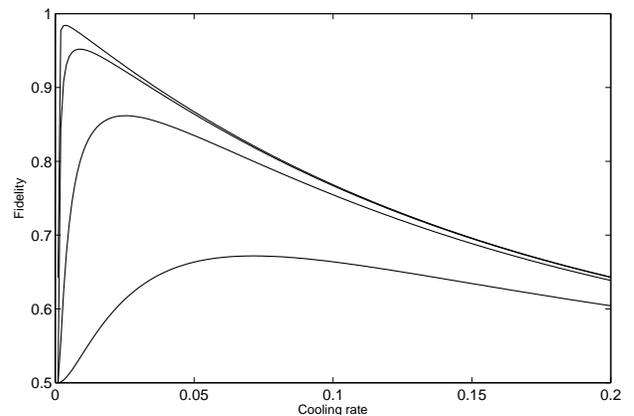}
  \caption{Fidelity as a function of the cooling constant $\alpha_{\mbox{\tiny bath}}$ for couplings to the background (from the bottom to the top) $\gamma = 10^{-2}, 10^{-3}, 10^{-4}$, and $10^{-5}$. For each there is an optimal cooling rate that maximizes the fidelity.}\label{fidelity2}
\end{figure}

To understand the effect of cooling on the fidelity of a computation, we consider the three qubit example of Sec. \ref{subsec:Xrot}, using the master equation \eqref{twobaths}. We assume that the system is initially in equilibrium under
(\ref{twobaths}) and that the measurements are performed on qubits 1
and 2 at times $t=0$ and $t=2\pi /\Delta$. Between the measurements
the system evolves according to the master equation
\eqref{twobaths}, after which we calculate the fidelity between the
actual output state on qubit 3 and the ideal output state. The
behaviour of the fidelity as a function of the coupling $\alpha_{\mbox{\tiny bath}}$, for various values of  $\gamma$, is
shown in Fig.~\ref{fidelity2}.

From Fig. ~\ref{fidelity2} it can be seen that, for large $\alpha_{\mbox{\tiny bath}}$, there is a high loss in fidelity due to
rapid dynamics in between the measurement steps, that try to bring
the system back to equilibrium. At the other extreme, for a weak coupling to the cooling bath $\alpha_{\mbox{\tiny bath}} \to 0$, the initial state of the system is highly mixed, due to the coupling to the warm background, and so the fidelity is also reduced. There is consequently a trade-off in terms of cooling strength,
between counteracting the heating effects of the background and
reducing errors due to dynamics between measurements for MBQC on the system. Thus given any
ambient background there is an optimal coupling of the system to
the cooling bath. Provided the coupling to this cooling bath is under the control of the experimentalist, the optimal coupling should be selected in order to maximise the fidelity of computations.

Note that the two Lindblad terms in \eqref{twobaths} can be absorbed into a single term, such that the effect of the two baths is the same as coupling to a single bath with $\alpha = \alpha_{\mbox{\tiny bath}}  + \gamma$ and $\beta = \gamma$, so that using (\ref{temperature}), the effective temperature of the bath is given by $kT = \Delta / \log(1+ \alpha_{\mbox{\tiny bath}} /\gamma) $. In the subsequent sections we treat the system as if it were coupled to a single bath parameterized by $\alpha$ and $\beta$.

\section{General Decoherence in MBQC}
\label{sec:General}

For our simple $X$-rotation gate on a 3 qubit state, we found that
the loss in fidelity of the encoded qubit takes on a particularly
simple form. In this section, we generalize this result for an
arbitrary sequence of measurements in a MBQC scheme, performed at
the multiples of the natural timescale $\tau$.  Within a general
MBQC scheme on a $d$-dimensional lattice, one dimension is
identified as ``time'' and a $(d-1)$-dimensional \emph{logical}
state evolves through the lattice via measurements.  We show that
the decoherence of this logical state as MBQC proceeds along the
time direction coupled via $H_I$ to a bath at a given temperature is
described by a single quantum operation, acting on the logical
state, producing anisotropic decoherence towards the maximally mixed
state. The importance of this result is that the error model for the
logical qubit is Markovian when we restrict to measurements at
multiples of $\tau$ on the cluster state.  This error model in turn
allows for the application of standard fault-tolerant thresholds.

\subsection{One-dimensional lattice}

We begin by considering single-qubit unitaries through MBQC on
one-dimensional lattices, and will consider the general case in the
next section.  On a line, with qubits labeled sequentially left to
right, consider the situation of already having performed $N-1$
projective measurements after a time $(N-1)\tau $, where $\tau
=2\pi/\Delta$ is the natural measurement time. Consequently, the
qubit at site $N-1$ is in a pure state, while the qubits $i<N-1$ are
partially entangled having evolved back towards the cluster state
under the full Hamiltonian for the spin-lattice system coupled to
the thermal bath. The qubits $i>N-1$ are in an entangled state and
are still awaiting measurement.
\begin{center}
\includegraphics[width=7.5cm ]{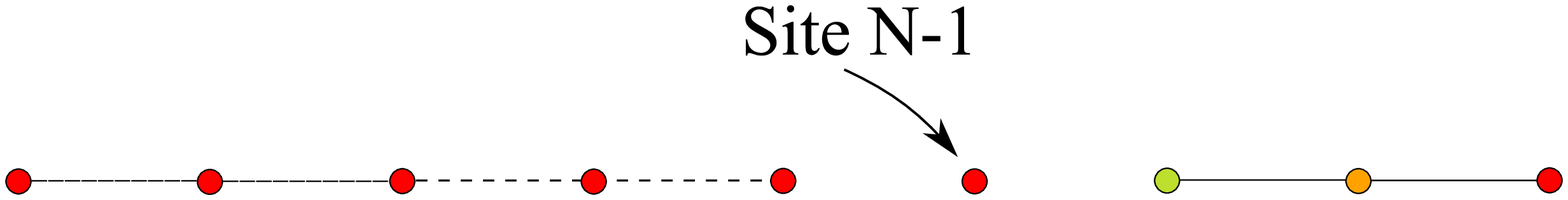}
\end{center}

We map the qubits $N,N+1,N+2, \cdots$ to a system of unentangled
qubits by applying the unitary
\begin{equation}
  V_N = \prod_{i>N} U_{i-1,i}\,,
\end{equation}
where $U_{i-1,i}$ is the controlled-$Z$ unitary between qubit $i-1$
and qubit $i$. This map transforms the Hamiltonian $H_{\mbox{\tiny
tot}} \rightarrow H'_{\mbox{\tiny tot}} = V_N H_{\mbox{\tiny tot}}
V_N^\dagger$ such that qubits $i>N-1$ are uncoupled, and more
importantly localizes the logical state to qubit $N$,

\begin{center}
\includegraphics[width=7.5cm ]{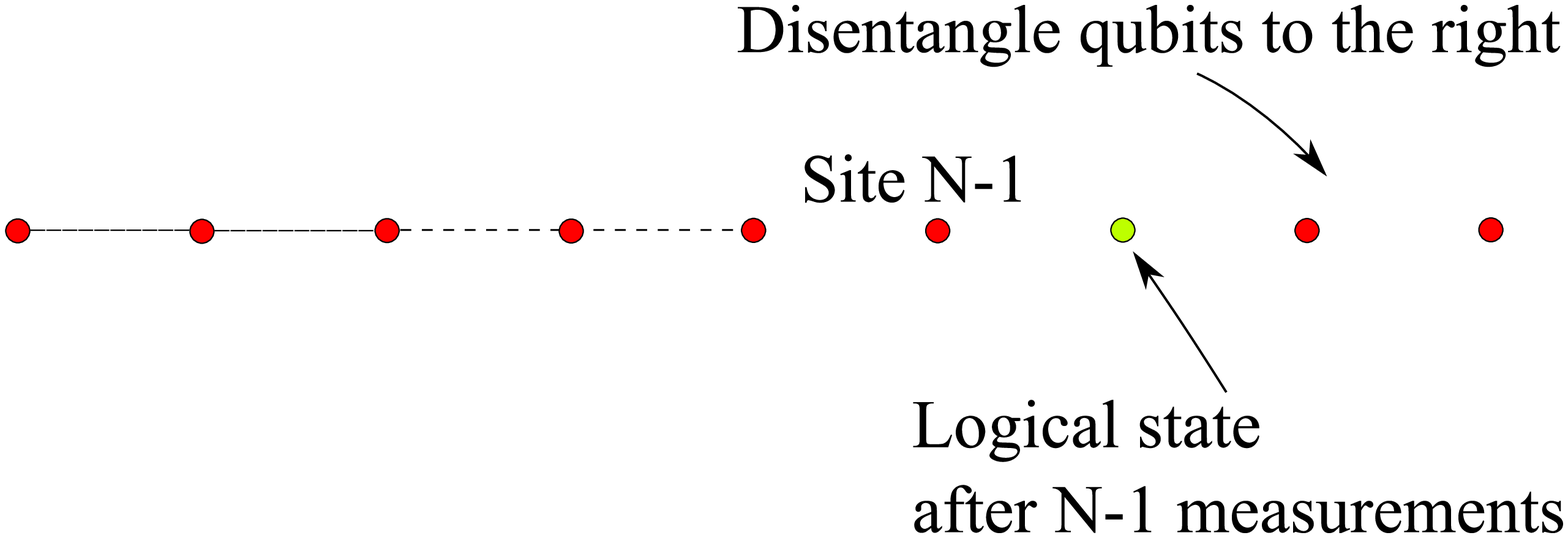}
\end{center}
The dynamics of the original state is determined by mapping under
$V_N$, evolving under $H'_{\mbox{\tiny tot}}$ and then mapping back
with $V_N^\dag$. However, because the qubits at sites $i>N$ are
assumed to be in equilibrium, they are static under the dynamics and
may be ignored, and so we only have to consider the dynamics of the
measured qubits together with the logical state at qubit $N$. The
logical state, localized at site $N$, decoheres during the time
interval $(N-1) \tau $ to $N \tau $ under $H'$, where the qubit
interacts with site $N-1$ through the cluster Hamiltonian term
$-\frac{\Delta }{2} Z_{N-1} X_N$, and with the bath through the
$\sum_j \lambda_{Nj} Z_N q_{Nj}$.

The state of the measured qubits and qubit $N$ evolves according to the map $\E^{\mbox{\tiny
tot}}_\tau (\rho )= \sum_I W_{I,\tau} \rho W_{I,\tau} ^\dagger $,
where the Kraus operators $W_I$ for the entire system can be
expressed in terms of the Kraus operators $M_i$ of Eq.~(\ref{kraus})
as
\begin{equation}
 W_{I=\{ i_1, \dots,i_N \},\tau } =  U (M_{i_1,\tau} \otimes M_{i_2,\tau} \otimes \cdots \otimes M_{i_N,\tau} )U^\dagger
\end{equation}
and where $U$ is the product of all the controlled-$Z$
unitaries for bonds to the left of site $N$.

The evolution has the effect of partially entangling the logical
state at $N$ with the other qubits.  The input state for the
$N^{\mbox{\tiny th}}$ projective measurement is then
\begin{equation}
  \Tr_{1,\dots,N-1} [ \E^{\mbox{\tiny tot}}_\tau  (\rho)]  =
  \frac{1}{2} (\I + \vec{r}_{N \tau} \cdot \vec{\sigma} )\,,
\end{equation}
and we obtain the components of its Bloch vector $\vec{r}_{N \tau}$
via $\vec{r}_{N\tau} = \Tr [\vec{\sigma  } \E^{\mbox{\tiny
tot}}_\tau (\rho) ]$.  Consequently, we can determine the equations
of motion for this vector as a function of time from the master
equation.

For simplicity, we go to the interaction picture, with $\rho_I (t) = \exp
[-iH_c t ] \rho(t) \exp [iH_c t ] $ and obtain the equations
\begin{align}\label{eqnmotion}
 \partial _t \<X_N (t) \>_I &= (\alpha - \beta ) \< Z_{N-1}(t) \>_I - \frac{3 }{2} (\alpha + \beta ) \<X_N (t)\>_I \non
 \partial _t \<Y_N (t) \>_I &= - (\alpha + \beta ) \<Y_N (t)\>_I \non
 \partial _t \<Z_N (t) \>_I &= - \frac{1}{2}(\alpha + \beta ) \<Z_N
 (t)\>_I\,,
\end{align}
where $\< A(t) \>_I = \Tr [ \rho_I (t) A]$ for any observable $A$
and $\alpha $ and $\beta $ the temperature dependent coupling
parameters as in (\ref{master}). The equation for the $Z$ component
holds for any site $s$ and so $\< Z_s (t) \>_I = z_s \exp
[-\frac{1}{2} (\alpha + \beta )t ]$. Now all computational
measurements in MBQC on the cluster take place in the $X$-$Y$ plane,
and so $\<Z_s(t) \> = 0$ for all $s<N$ and for all time $t$ after
site $s$ has been measured.

The components of the logical state in the interaction picture then
evolve as
\begin{align}
  \<X_N (t)\>_I &= x_N \exp [-\frac{3}{2}(\alpha + \beta )t ]\,,
  \nonumber \\
  \<Y_N (t)\>_I &= y_N \exp [-(\alpha + \beta )t ]\,, \nonumber \\
  \<Z_N (t)\>_I &= z_N \exp [-\frac{1}{2}(\alpha + \beta )t  ]\,.
\end{align}
However, at times $t=n\tau$, we have $\rho _I (t) = \rho (t)$, and
so for these times the decoherence to the maximally mixed state is
deduced from the interaction picture results, and agrees with the
explicit example of the 3 qubit system in Sec.~\ref{subsec:Xrot}.

The result of this analysis is that the MBQC scheme along the line
of qubits with a free Hamiltonian and in contact with a bath at a
finite temperature can be described in simple terms for any sequence
of measurements on the logical state at times which are multiples of
$\tau$, using a fixed Markovian noise operator $\mathcal{F}$.  For a
sequence of measurements $\{ \phi _1, \phi _2, \dots \phi _N\}$ in
the $X$-$Y$ plane labelled at each site by an angle $\phi $ from the
$X$ axis and with outcomes $\{s_1,s_2, \dots, s_N \}$, the single
qubit logical state is processed as
\begin{align}
\psi_{\mbox{\tiny in}} & \rightarrow [Z^{s_1} X(\phi _1)H]
[\psi_{\mbox{\tiny in}}]\rightarrow \F[[Z^{s_1} X(\phi _1)H]
[\psi_{\mbox{\tiny in}}]]\rightarrow \nonumber \\
& \rightarrow  [Z^{s_2} X(\phi _2)H] [\F[[Z^{s_1}X(\phi _1)H]
[\psi_{\mbox{\tiny in}}]]]\rightarrow \nonumber \\
& \rightarrow \F[ [Z^{s_2}X(\phi _2)H] [\F[[Z^{s_1}X(\phi _1)H]
[\psi_{\mbox{\tiny in}}]]]]\rightarrow \cdots \,,
\end{align}
where for any $A$ we denote $[A][\rho ] =A \rho A^\dagger$, $H = |+
\> \< 0 | + |- \> \< 1|$, $X(\phi) = \exp (-i\frac{\phi}{2}X)$, and
$\F [\rho ]$ is the quantum operation given by
\begin{equation}\label{qop}
  \F [\rho ] =p_1 \rho + p_2  X\rho X + p_3 Y \rho Y + p_4 Z \rho
  Z\,,
\end{equation}
with
\begin{align}
 p_1 &=\frac{1}{4} (1+w)(1+w^2) \hspace{0.5cm} p_2=\frac{1}{4}(1-w)(1-w^2) \non
  p_3 &= \frac{1}{4}  (1-w)(1+w^2) \hspace{0.5cm}  p_4 =\frac{1}{4} (1+w)(1-w^2)
\end{align}
and $w=\exp [-(\alpha + \beta ) t/2 ]$.

It is also clear from this analysis why in the case of an arbitrary $X$
rotation performed with three qubits, that the first evolution is
slightly different from the second one: before the first measurement
there are no qubits to the left of the first site to affect the
logical state, while the qubits to the right are already in their
equilibrium state, and so the evolution of the state is localized to
the first site.

\subsection{General lattices}

The analysis of the last section can be extended to
higher-dimensional lattices. For example, if upon localization to a
site $s$ using an analogous unitary to $V_N$, the single qubit logical
state has $k$ neighbouring sites, labeled $1,2,3, \dots, k$, then
(\ref{eqnmotion}) generalizes to
\begin{align}\label{eqnmotion2D}
 \partial _t \<X_s (t) \>_I &= (\alpha - \beta ) \< Z_1 \cdots Z_k (t) \>_I  \non
&\quad - \frac{k+2 }{2} (\alpha + \beta ) \<X_s (t)\>_I \non
 \partial _t \<Y_s (t) \>_I &= - \frac{k+1}{2}(\alpha + \beta ) \<Y_s (t)\>_I \non
 \partial _t \<Z_s (t) \>_I &= - \frac{1}{2}(\alpha + \beta ) \<Z_s (t)\>_I \non
 \partial _t \<Z_1 \cdots Z_k (t) \>_I &= - \frac{k}{2}(\alpha + \beta ) \<Z_1\cdots Z_k (t)\>_I \,.
\end{align}

For general MBQC on for example the 2D or 3D lattice, we assume that
the adaptive measurements are performed in steps, with a time
interval $\tau$ before the next round of measurements. Each round of
measurements is composed of a set of $Z$ measurements to eliminate
qubits from the lattice and a set of measurements in the $X$-$Y$
plane to propagate correlations and to perform the desired
computational transformations.

After $N-1$ such measurement steps, we may formally disentangle the qubits to be measured at
steps $N, N+1,N+2,\dots$ and transform to a system where the logical state is localized to a set of qubits, which we denote
$\mathcal{Q}_N$. Once again, the qubits to measured at stages $N+1, N+2, \dots$ are in their equilibrium states and can be ignored for the timestep. The reduced
state for the logical state after a time $\tau$ is determined from terms of the form $\<M\>_I$, where $M$ is a
product of Pauli operators on $\mathcal{Q}_N$ and its surrounding
qubits.
\begin{figure}[t]
\centering
\includegraphics[width=7cm]{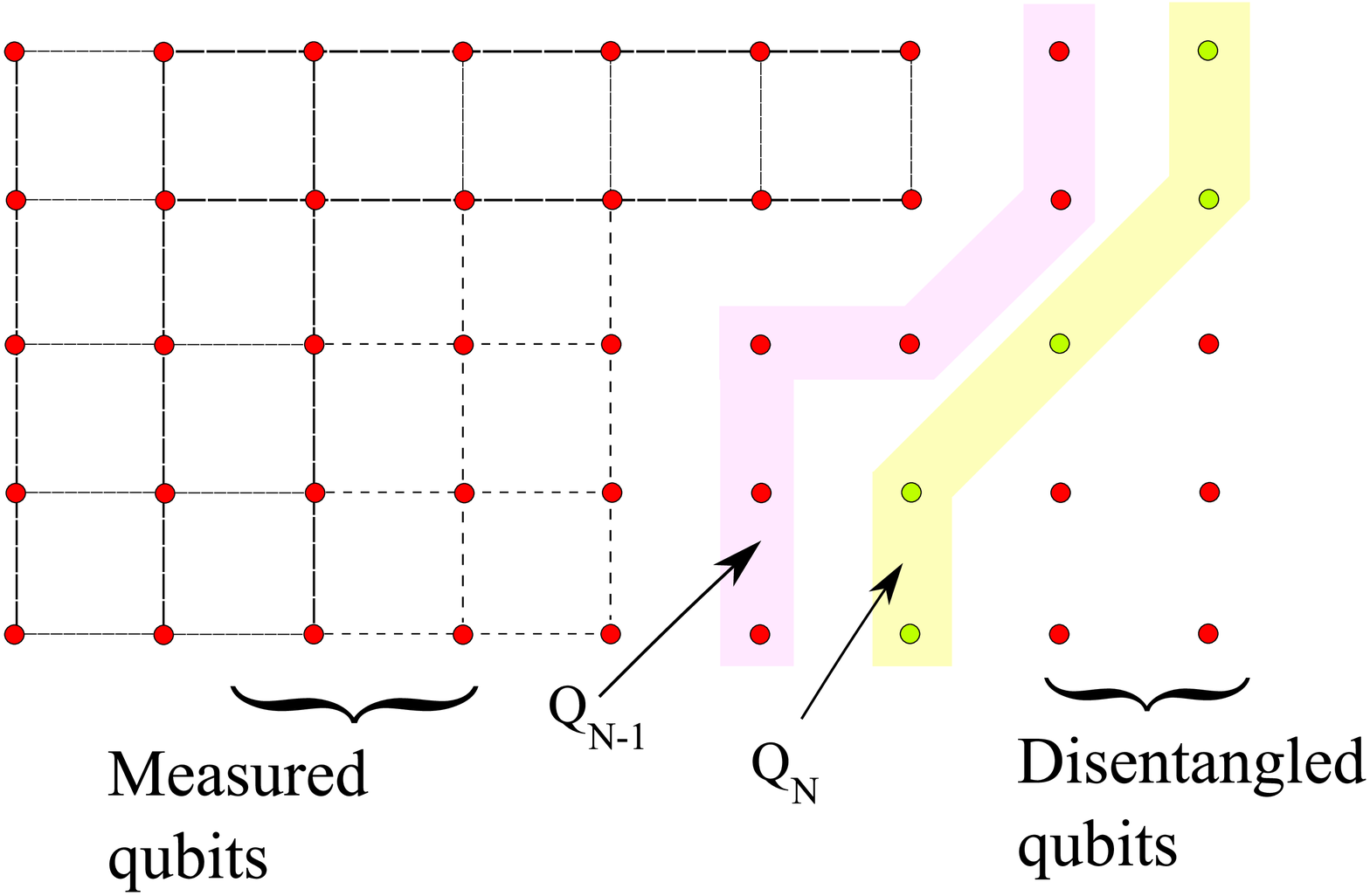}
\caption{Schematic diagram for the localization of the logical state
to the qubits  $\mathcal{Q}_N$ after $N-1$ timesteps. The qubits to
the right of $\mathcal{Q}_N$ are disentangled with a unitary $V_N$,
while those to the left have already been measured. The qubits
$\mathcal{Q}_{N-1}$ are in a pure product state, having just been
measured. } \label{fig:}
\end{figure}

A site $i$ on the lattice contributes to the equation of motion of a
general observable $M$ according to the following rule: it
contributes $-\frac{(\alpha+\beta)}{2}\<M\>_I$ if either
$[M,Z_i]=\{M,K_i\}=0$ or if $\{M,Z_i \}=\{M,K_i \}=0$; it
contributes $-(\alpha+\beta)\<M\>_I + (\alpha-\beta) \<MK_i\>_I$ if
$\{M,Z_i \} =[M,K_i]=0$; and zero otherwise. Consequently, awkward
terms can arise when $M$ contains an $X$ observable. These terms
couple the equations of motion for observables on $\mathcal{Q}_N$
with observables on $\mathcal{Q}_N$ plus its neighbours, however for
MBQC on the cluster state these equations decouple, from the
following argument.

For an observable $M$ on $\mathcal{Q}_N $ containing $m$ $X_i$
observables, its equation of motion will only have terms of the form
$\<M\>_I$ and $\<M K_i\>_I$. However $MK_i$ is an observable with
$(m-1)$ $X$'s on $\mathcal{Q}_N$ and a number of $Z$'s in
$\mathcal{Q}_{N-1}$, the set of qubits that have just been measured.
If we iterate and obtain the full set of coupled equations that
determine $\<MK_i\>_I$ we arrive at a dependence on observables
without any $X$ observables and with at least one $Z_s$ on site $s$
in $\mathcal{Q}_{N-1}$. The equation of motion for such an
observable $M'$ is of the form $\partial_t \<M'\>_I = -p(\alpha
+\beta)/2 \<M'\>_I$ for some integer $p$. Furthermore, if $s$ was
measured in the $X$-$Y$ plane then initially $\<M'\>_I=0$ and so
will remain zero for the whole time interval. Retracing the chain of
coupled equations we find that each problematic term of the form
$\<MK_i\>_I$ vanishes for $t\ge 0$ and the equations of motion for
the observables on $\mathcal{Q}_N$ are decoupled provided each qubit
in $\mathcal{Q}_N$ has at least one neighbour in $\mathcal{Q}_{N-1}$
that was measured in the $X$-$Y$ plane.

The expectation of an observable $M$ on $\mathcal{Q}_N$ will evolve
as $\<M\>_I = M_0 e^{-q(\alpha+\beta)t/2}$ for some integer $q$, and
for Pauli observables $M_1, M_2 \dots ,M_k$ on sites $1,2, \dots ,k$
we have that $|\<M_1\cdots M_k(t)\>_I | \ge| \<M_1(t)\>_I| \cdots
|\<M_k(t)\>_I|$, with equality coming when the sites do not share
any neighbours.

\subsection{Fault tolerance}

With a simple Markovian description of the errors present in our
scheme, we can consider fault-tolerant MBQC.  There are two sources
of errors in the dynamical setting that we are considering. First,
the equilibrium state for the system is at a non-zero temperature,
and so there are preparation errors due to an imperfectly prepared
cluster state.  Second, errors occur due to the dynamics between
measurements, and can be viewed as storage errors on the qubits for
a given timestep. For sufficiently low rates, MBQC on a 3D lattice
has been shown to be fault-tolerant for both of these sources of
errors~\cite{Rau06,Rau07}. If the state distillation protocol of
Ref. ~\cite{Rau07} is used, the error threshold is set by the bulk
topological part of  the error correction scheme, which in turn can
be related to a phase transition in the classical random plaquette
gauge model ~\cite{Wang02}.


Our initial state is a thermal state, static under the dynamics, prepared by cooling
with the bath. Such a thermal cluster state at a temperature $T$ is
obtained by applying $Z$ errors to a perfect cluster state with
probability $p_{\mbox{\tiny prep}}= (1+\exp (\Delta /(kT)))^{-1}$.


For the cubic lattice model, the dynamics in between measurement
steps produce an error channel on the individual qubits no worse
than a quantum operation of the same form as (\ref{qop}) but with
coefficients
\begin{align}
 p_1 &=\frac{1}{4} (1+w)(1+w^6) \hspace{0.5cm} p_2=\frac{1}{4}(1-w)(1-w^6) \non
  p_3 &= \frac{1}{4}  (1-w)(1+w^6) \hspace{0.5cm}  p_4 =\frac{1}{4} (1+w)(1-w^6)\, ,
\end{align}
and with $w=e^{-(\alpha+\beta)\tau /2}$. Consequently, the resultant
errors for a cubic lattice are no worse than those obtained by
application of the local depolarizing channel $T[\rho] = (1-p_s)
\rho + \frac{p_s}{3} (X\rho X +Y\rho Y +Z \rho Z)$ with
$p_s=\frac{1}{4}(1+w)(1-w^6)$, on each individual qubit.

The combined effect of these two errors leads to independent errors
on each qubit in the lattice with effective parameter $q =
p_{\mbox{\tiny prep}} + \frac{2}{3}  p_s$ (c.f. ~\cite{Rau06}). The
threshold for such errors is given by $q < 0.0293$  ~\cite{Wang02}.
Thus, if errors due to the dynamics can be neglected, i.e. when $
p_s \to 0 $, the error threshold for preparation errors corresponds
to a temperature bound of $T\sim 0.28 \Delta$. Conversely, if errors
due to preparation can be neglected, $ p_{\mbox{\tiny prep}} \to 0$,
the error threshold corresponds to a threshold for the environmental
couplings of $(\alpha+\beta)/ \Delta \sim 4.6 \times 10^{-3}$.

If this environment consists of a infinite temperature background
parametrized by $\gamma$ and a zero-temperature cooling bath
parametrized by $\alpha_{\rm bath}$ as in Sec.~\ref{cooling}, the
parameter $q$ is a function of these two parameters. The constraint
$q(\alpha_{\rm bath}, \gamma) =0.0293$ defines the threshold value
of $\gamma$ implicitly in terms of $\alpha_{\rm bath}$.  We may then
maximize this $\gamma$ over the bath couplings and deduce an overall
threshold of $\gamma/\Delta \sim 3.4 \times 10^{-5}$ for the
coupling to the environment provided that the cooling rate for the
bath is set at a ``Goldilocks value" of $\alpha_{\rm bath} /\Delta
\sim 2.26 \times 10^{-3}$. For this cooling rate the system is not
so hot that large preparation errors destroy entanglement, and it is
not so cold that large storage errors erase the logical state.  That
is, if the coupling to the background environment $\gamma$ is below
this threshold, it is possible to devise a cooling bath that allows
for fault-tolerant MBQC.

\section{Discussion}

When the ground state of a physical system provides a resource state
for MBQC we must necessarily take into account the system's
dynamics. As we have discussed there are several sources of
complication compared with MBQC on a static resource. While we may
prepare the system very close to its ground state, any measurements
we then perform on it will produce excitations and for a general
adaptive measurement scheme, involving classical feed-forward, the
resultant dynamics between measurements will perturb the state and
affect the computation.

For the simple, dispersionless Hamiltonian (\ref{hamc}) describing a
lattice of spins we showed that measurements should be performed at
a characteristic clock speed $2 \pi / \Delta$ defined via the energy
gap $\Delta$. However, the presence of environmental interactions
further complicates matters. For the environment, we considered both
ambient background effects and also the effects of a thermal bath
used to prepare and maintain the lattice system. We found that an
optimal cooling exists, which is a trade-off between adequate
shielding of the system from a hot background and providing slow
dynamics that allow adaptive measurements. Furthermore, the loss in
fidelity due to this dynamics is conveniently described in terms of
a single quantum operation (\ref{qop}) that acts on the logical
state.

The importance of our results is that under certain conditions, the
environment produces Markovian errors on the logical state and is
thus amenable to error-correction. Our results are general and do
not depend on the type of lattice or its dimensionality. In the
particular case of a cubic lattice we may invoke fault-tolerance
results for MBQC in the presence of local independent depolarizing
errors to obtain a threshold of $T \sim 0.28 \Delta$ for the
temperature of the prepared state when dynamics may be neglected,
and a threshold of $(\alpha+ \beta)/ \Delta \sim 4.6 \times 10^{-3}$
for the ratio of environmental couplings to energy gap when the
storage errors dominate. In addition, we obtained a threshold of
$\gamma/\Delta \sim 3.4 \times 10^{-5}$ for the coupling to a high
temperature environment provided there is a zero-temperature cooling
bath with coupling $\alpha_{\rm bath} /\Delta \sim 2.26 \times
10^{-3}$ to the lattice system.

Several issues remain that deserve investigation. For example, we
have not discussed possible imperfections in the Hamiltonian or
measurement errors, both of which would modify the above thresholds.
Furthermore, the free Hamiltonian behaviour suggests the obvious
strategy of performing all measurements at or near the clock cycles
of $\tau=2 \pi /\Delta$. However more complicated measurement
strategies may exist that produce high fidelities in the presence of
a fixed cooling.

While the above formalism may be adapted to different
settings or more particular questions, another key outstanding issue
is the effect of finite-time measurements in which the measurements
themselves are not instantaneous but are spread over some small
finite interval of time. Such a situation requires a more elaborate
analysis than the one presented here, especially when the
measurement time becomes comparable with the clock cycle time
$\tau$.

\begin{acknowledgments}
We thank Chris Dawson for early discussions. D.J.\ and T.R.\ acknowledge the support of the EPSRC.  S.~D.~Barrett acknowledges the support of the EPSRC and the Centre for Quantum Computer Technology.  S.~D.~Bartlett acknowledges the support of the Australian Research Council.
\end{acknowledgments}

\end{document}